\documentclass[aps,prd,preprintnumbers,superscriptaddress,nofootinbib,notitlepage,floatfix,10pt]{revtex4-2}
\usepackage[pdftex]{graphicx}
\usepackage{bm,latexsym,amsmath,amssymb,amsfonts,mathrsfs}
\usepackage{physics}
\usepackage{picture}
\usepackage{here}
\usepackage{color}
\allowdisplaybreaks[1]
\usepackage[pdftex,colorlinks=true,linkcolor=blue,citecolor=cyan,backref=page]{hyperref}
\newcommand*{\D}{\mathrm{d}}

\newcommand{\pd}{\partial}
\newcommand{\cd}{\nabla}
\newcommand{\mrm}[1]{\mathrm{#1}}

\newcommand{\mcal}[1]{\mathcal{#1}}

\def\om{\Omega_{\mathrm{m}}}
\def\omp{\Omega_{\mathrm{m}0}}

\begin{document}
\title{Abundance of cosmic voids in EFT of dark energy}
%
\author{Toshiki~Takadera}
\email[Email: ]{t\_takadera@rikkyo.ac.jp}
\affiliation{Department of Physics, Rikkyo University, Toshima, Tokyo 171-8501, Japan}
\author{Shin'ichi~Hirano}
\email[Email: ]{shirano@hyogo-u.ac.jp}
\affiliation{Hyogo University of Teacher Education, 942-1 Shimokume, Kato, Hyogo 673-1494, Japan}
\affiliation{Department of Physics, Rikkyo University, Toshima, Tokyo 171-8501, Japan}
\author{Tsutomu~Kobayashi}
\email[Email: ]{tsutomu@rikkyo.ac.jp}
\affiliation{Department of Physics, Rikkyo University, Toshima, Tokyo 171-8501, Japan}
%
\begin{abstract}
Cosmic voids in the large-scale structure are among the useful probes for testing gravity on cosmological scales.
In this paper, we investigate the evolution of voids in the Horndeski theory using the effective field theory (EFT) of dark energy.
Modeling the void formation with the dynamics of spherical mass shells, we study how modifications of gravity encoded into the EFT of dark energy change the linearly extrapolated critical density contrast that is relevant for the criterion for void formation, with particular focus on the time-dependent parameter characterizing the effect of kinetic braiding.
It is found that the change in the critical density contrast is one order of magnitude smaller than the dimensionless EFT parameter because of a slight imbalance between two compensating effects.
We then compute the void abundance using the Sheth--van de Weygaert void size function and demonstrate that it exhibits scale-dependent modifications.
It is shown that the modifications to the void size function on small scales are almost entirely determined by the modified linear matter power spectrum, while the modifications on large scales are dominated by the contributions from the linear matter spectrum and the critical density contrast.
\end{abstract}

\preprint{RUP-25-26}
\maketitle

\section{Introduction}\label{sec:intro}

The discovery of the accelerated expansion of the Universe through observations of type Ia supernovae~\cite{SupernovaCosmologyProject:1998vns,SupernovaSearchTeam:1998fmf} marked a major milestone in modern cosmology.
The current standard framework for describing this accelerated expansion is the $\Lambda$CDM model, which assumes general relativity as the underlying theory of gravity and introduces a positive cosmological constant $\Lambda$ to drive the late-time acceleration.
This model successfully accounts for a wide range of cosmological observations with only a small number of parameters.
However, it also faces several challenges.
From a theoretical perspective, explaining the tiny value of $\Lambda$ requires an extreme degree of fine-tuning.
From an observational perspective, there remain persistent, albeit mild, tensions between $\Lambda$CDM predictions and various cosmological datasets.
These issues motivate the exploration of alternative explanations for cosmic acceleration.
One possible direction is to consider modifications of gravity on cosmological scales.

Scalar-tensor theories constitute not only the simplest but also the most extensively studied class of modified gravity models, and can effectively describe a wide variety of modified gravity scenarios.
In scalar-tensor theories, general relativity is extended by introducing an additional scalar degree of freedom on top of the two tensor degrees of freedom corresponding to gravitational waves.
The Horndeski theory is the most general scalar-tensor theory with second-order field equations~\cite{Horndeski:1974wa,Deffayet:2011gz,Kobayashi:2011nu},
and can describe each specific model of modified gravity when the functions in the action are specified.
See Ref.~\cite{Kobayashi:2019hrl} for a review.
It has been applied extensively to the study of cosmology in modified gravity (see, e.g., Ref.~\cite{Arai:2022zzz}).

In the Horndeski family of theories, the characteristic screening mechanism, known as the Vainshtein mechanism, can operate due to nonlinear derivative interactions of the scalar degree of freedom, restoring the gravitational field predicted in general relativity in the vicinity of a source~\cite{Kimura:2011dc,Narikawa:2013pjr,Koyama:2013paa}.
In the context of cosmology, the operation of this mechanism relies on the nonlinearity of matter density perturbations.
In this paper, we focus on the large-scale structure (LSS) of the Universe, where the nonlinearity of matter density perturbations must be taken care of appropriately.

The LSS of the Universe provides one of the most powerful observational probes for testing gravity on cosmological scales.
It is composed of several elements, including halos, filaments, walls, and voids.
Among these, we focus on cosmic voids, vast underdense regions that occupy most of the volume of the Universe.
In contrast to halos, the density contrast $\delta$ in voids has the limit $\delta \ge -1$ by definition, and hence $|\delta|$ cannot be much larger than one.
As a consequence, the Vainshtein screening mechanism is expected to work only partially in void environments, potentially leaving some signatures of modified gravity.
Indeed, Ref.~\cite{Falck:2014jwa} demonstrated that particles inside voids feel unscreened scalar-mediated forces in the nDGP model, a modified gravity model that can be described effectively by a subset of the Horndeski family.
For this reason, it is possible that cosmic voids serve as excellent laboratories for testing gravity on cosmological scales.

Several studies have investigated the properties of cosmic voids in modified gravity.
The authors of Ref.~\cite{2013MNRAS.431..749C} examined the evolution of voids in chameleon theories using an extended spherical collapse model based on the dynamics of spherical shells.
In Ref.~\cite{Voivodic:2016kog}, the abundance of cosmic voids was computed and compared with results from $N$-body simulations in $f(R)$ gravity and symmetron models.
Building on simulation-based approaches, Ref.~\cite{2012MNRAS.421.3481L} analyzed the void population in $f(R)$ gravity with $N$-body simulations and computed the void number density as a function of volume.
In the context of the nDGP model, Ref.~\cite{Falck:2017rvl} explored the impact of modified gravity on the density, velocity, and screening profiles of voids.
A similar analysis was performed in Ref.~\cite{Perico:2019obq} for $f(R)$ gravity and symmetron models (see also Refs.~\cite{Wilson:2022ets,Winther:2015pta,Zivick:2014uva,Contarini:2020fdu,Cai:2014fma,Mauland:2023eax}).
Weak-lensing signatures of voids were investigated in the nDGP model~\cite{Paillas:2018wxs,Davies:2019yif} and in the Galileon theories~\cite{Su:2022yoj,Baker:2018mnu,Barreira:2015vra}.

The studies mentioned above rely on specific models of modified gravity, each of which constitutes a subset of the Horndeski family.
To investigate gravity theories in a more model-independent manner, in this work we employ the effective field theory (EFT) of dark energy, a general and unified framework for studying cosmology in scalar-tensor theories~\cite{Gubitosi:2012hu,Gleyzes:2013ooa,Bloomfield:2012ff,Bellini:2014fua,Cusin:2017mzw}. 
This approach enables us to analyze modifications of gravity without specifying the explicit functional forms of the free functions in the Horndeski action, while assuming the background evolution and the time dependence of the coefficient of each term in the effective action.

In this paper, we model the void formation as the evolution of spherical mass shells.
In this simplified modeling, the criterion for the formation of a void is given by the intersection of two neighboring shells at the outer boundary of the underdense region.
One can then assess how the EFT of dark energy parameters change the linearly extrapolated critical density contrast for the void formation.
Combining this with several other quantities affected by gravity modification, we calculate the void size function (VSF)~\cite{Sheth:2003py}, i.e., the number density of voids as a function of their volume, and discuss the impacts of the EFT of dark energy parameters on it.

This paper is organized as follows.
In Sec.~\ref{sec:pre}, we present the basic equations governing the mass shell dynamics and the gravitational fields in the EFT of dark energy.
In Sec.~\ref{sec:void}, we analyze the evolution of an underdense region, and examine how the time-dependent EFT functions change the critical density contrasts for the void formation.
We then compute in Sec.~\ref{sec:vsf} the VSF using the Sheth--van de Weygaert model and discuss the impact of modified gravity on it.
Finally, we summarize our findings in Sec.~\ref{sec:conclusions}.

\section{Preliminaries}\label{sec:pre}

As preliminaries for the subsequent sections, we first present the equation of motion for a spherical shell composed of nonrelativistic matter, which can be used for the study of the evolution of an underdense region.
We then introduce the EFT of dark energy that defines modified gravity theories considered in this paper, as well as the cosmological background model we will assume and the gravitational field equations derived from the EFT action.

\subsection{The basic equation of a shell for spherical underdensity}
\label{subsubsec:shell}

We consider the metric in the Newtonian gauge,
\begin{align}
    ds^2=-\left[1+2\Phi(t,\vec{x})\right]dt^2+a^2(t)\left[1-2\Psi(t,\vec{x})\right]d\vec{x}^2.
    \label{eq:Newtonian}
\end{align}
We consider a spherically symmetric configuration of a nonrelativistic fluid described by the density $\rho(t,R)$ and the radial velocity $u(t,R)$, where we use the physical radius $R$ rather than the comoving radius and $u$ includes the Hubble flow $HR$, where $H(t)=\dot{a}/a$ and a dot denotes $\D/\D t$.
The continuity and Euler equations for the fluid are given, respectively, by
\begin{align}
    \frac{\partial \rho}{\partial t}+\frac{1}{R^2}\frac{\partial}{\partial R}
    \left(R^2\rho u\right)&=0,\label{eq:fluid:condinuity}
    \\ 
    \frac{\partial u}{\partial t}+u\frac{\partial u}{\partial R}
    -\left(\dot H+H^2\right)R&=-\frac{\partial\Phi}{\partial R}.
    \label{eq:fluid:euler}
\end{align}

To study the evolution of a spherical underdense region, one usually adopts the formulation based on concentric spherical shells with physical radius $R$ at time $t$, centered on the underdense region, instead of using the above fluid equations directly.
Since the shells consist of nonrelativistic matter, the particles constituting them obey the geodesic equation,
\begin{align}
      \frac{\D^2x^\mu}{\D\tau^2}+\Gamma^\mu_{\alpha\beta}\frac{\D x^\alpha}{\D\tau}\frac{\D x^\beta}{\D\tau}=0,
\end{align}
where $\tau$ is the proper time of a particle.

By using the physical radius $R=ar$ in stead of the comoving radius $r=\sqrt{x^2+y^2+z^2}$, the metric~\eqref{eq:Newtonian} can be written as
\begin{align}
    \D s^2=-\left(1+2\Phi\right)\D t^2+\left(1-2\Psi\right)
    \left[\left(\D R-HR\D t\right)^2+R^2\D\Omega^2\right],
\end{align}
where $\D\Omega^2$ is the line element of the unit 2-sphere.
In the spherically symmetric case, $\Phi$ and $\Psi$ depend only on $t$ and $R$.
The radial motion $R(t)$ of the particle is governed by the $t$ and $R$ components of the geodesic equations.
The time derivatives along the geodesic are obtained from $\D R/\D t=(\D R/\D \tau)/(\D t/\D \tau)$ and $\D^2 R/\D t^2=(\D^2 R/\D \tau^2-\dot R\D^2t/\D\tau^2)/(\D t/\D \tau)^2$, yielding
\begin{align}
    \frac{\ddot R}{R}&=\frac{\ddot a}{a}-\frac{1}{R}\frac{\partial \Phi}{\partial R}+\dots,
\end{align}
where the ellipsis indicates the terms of
$\mcal{O}(H^2R^2\cdot H^2)$, $\mcal{O}(H^2R^2\cdot R^{-1}\partial_R\Phi)$, $\mcal{O}(H\partial_t\Phi)$, $\mcal{O}(\dot R\cdot R^{-1}\partial_R\Phi)$, $\mcal{O}(\dot R^2H^2)$, and $\mcal{O}(\dot R^3\cdot H/R)$, which we will ignore relative to $\ddot a/a$ and $R^{-1}\partial_R\Phi$ by assuming that $HR\sim \dot R \ll 1$.
Now, by identifying $\dot R$ as $u(t,R(t))$, one can reproduce the Euler equation~\eqref{eq:fluid:euler} from the geodesic equation,
\begin{align}
    \frac{\ddot R}{R}=\frac{\ddot a}{a}-\frac{1}{R}\frac{\partial\Phi}{\partial R}.
    \label{eq:R}
\end{align}

Let us define a function $\mcal{M}(t,R)$ by
\begin{align}
    \partial_R\mcal{M}=4\pi\rho(t,R)R^2.\label{eq:dRM}
\end{align}
Then, it follows from the continuity equation~\eqref{eq:fluid:condinuity} that 
\begin{align}
    \partial_t\partial_R\mcal{M}=-4\pi\partial_R\left(R^2\rho u\right),
\end{align}
implying that
\begin{align}
    \partial_t\mcal{M}=-4\pi \rho R^2u.\label{eq:dtM}
\end{align}
Using Eqs.~\eqref{eq:dRM} and~\eqref{eq:dtM}, one can show that $\mcal{M}$ is conserved along the geodesic:
\begin{align}
    \frac{\D}{\D t}\mcal{M}(t,R(t))=\partial_t\mcal{M}+\dot R\partial_R\mcal{M}=0,
    \label{eq:mass-conservation}
\end{align}
where $u=\dot R$ was used.
Therefore, the value of $\mcal{M}$ for each geodesic is determined at some initial moment as
\begin{align}
    \mcal{M}(t,R(t))=\mcal{M}(t_i,R_i)=
    4\pi\bar\rho(t_i)\int_0^{R_i}\left[1+\delta(t_i,\tilde R)\right]\tilde R^2\D \tilde R,
\end{align}
with $R_i:=R(t_i)$.

So far in Sec.~\ref{subsubsec:shell}, we have not assumed any specific gravitational theory.
Equations~\eqref{eq:R} and~\eqref{eq:mass-conservation} hold as long as the nonrelativistic matter is minimally coupled to gravity and the particles obeys the geodesic equation.

\subsection{Gravitational field equations in EFT of dark energy}\label{subsec:basic_eq}

To study the impacts of modified gravity on void formation, we use the EFT of dark energy and modified gravity, which can be derived directly by expanding the Horndeski action in powers of the metric and scalar-field perturbations around a cosmological background.
We are interested in the subhorizon dynamics and are allowed to make the quasi-static approximation.
We keep the nonlinear derivative interaction terms in the EFT of dark energy, as they play an important role when the density contrast becomes of order unity and the Vainshtein mechanism starts to operate, even though the gravitational potential is much smaller than unity.

The effective action for the metric in the Newtonian gauge~Eq.\eqref{eq:Newtonian},
and for the scalar field, $\phi=t+\pi(t,\vec{x})$, is given by~\cite{Dima:2017pwp} (see also~\cite{Crisostomi:2017lbg,Langlois:2017dyl})
\begin{align}
        S=\int \D t\D^3x&\biggl\{\frac{M^2a}{2}\biggl[\left(c_1\Phi+c_2\Psi+c_3\pi\right)\cd^2\pi+c_4\Psi\cd^2\Phi+c_5\Psi\cd^2\Psi
    +\frac{b_1}{a^2}\mcal{L}^{\mrm{Gal}}_3+\frac{1}{a^2}(b_2\Phi+b_3\Psi)\mcal{E}^{\mrm{Gal}}_3\notag \\
  &+\frac{1}{a^4}d_1\mcal{L}^{\mrm{Gal}}_4\biggr]
  -a^3\Phi\delta\rho\biggr\},
    \label{eq:actionEFT}
\end{align}
where $\cd_i$ stands for the spatial derivative and
\begin{align}
    \mcal{L}^{\mrm{Gal}}_3&:=\frac{1}{2}(\cd\pi)^2(\cd^2\pi),\\
    \mcal{E}^{\mrm{Gal}}_3&:=(\cd^2\pi)^2-\cd_i\cd_j\pi\cd_i\cd_j\pi,\\
    \mcal{L}^{\mrm{Gal}}_4&:=\frac{1}{2}(\cd_i\pi)^2\mcal{E}^{\mrm{Gal}}_3.
\end{align}
Here, the matter inhomogeneity, $\delta\rho=\rho(t,\vec{x})-\bar{\rho}(t)=\bar{\rho}(t)\cdot\delta(t,\vec{x})$, is introduced, with $\bar{\rho}(t)$ being the mean matter energy density, where matter is assumed to be minimally coupled to gravity.
When expressed in terms of the time-dependent $\alpha$ functions used in the $\alpha$-basis EFT, the coefficients are given by
\begin{align}
    c_1&=2H\alpha_B,\quad c_2=4H(\alpha_M-\alpha_T),\notag \\
    c_3&=-\frac{\bar{\rho}}{M^2}+H\dot{\alpha}_B-\dot{H}(2-\alpha_B)+H^2[2\alpha_M+\alpha_B(1+\alpha_M)-2\alpha_T] ,\notag \\
    c_4&=4,
    \quad 
    c_5=-2(1+\alpha_T),
   \notag\\
    b_1&=H[-2\alpha_B-\alpha_V(1-\alpha_M)-2\alpha_M+3\alpha_T]+\dot{\alpha}_V,\quad b_2=\alpha_V,\quad b_3=\alpha_T,\notag \\
    d_1&=-(\alpha_T+\alpha_V).
\end{align}
These $\alpha$ functions and the effective Planck mass, $M$, can be related to the free functions in the Horndeski action,
and their explicit relations are presented in Appendix~\ref{app:EFT}.
We stress that the $b_1$, $b_2$, $b_3$, and $d_1$ terms are the nonlinear derivative interactions which trigger Vainshtein screening.
Note that the above effective action can be used even for $|\delta|\gtrsim 1$, provided that the gravitational potentials remain small, $|\Phi|,|\Psi|\ll1$.

If one starts from DHOST theories beyond Horndeski~\cite{Langlois:2015cwa}, more terms appear in the effective action, accompanied by additional $\alpha$ functions.
However, in this paper, we restrict ourselves to manifestly second-order scalar-tensor theories and use the EFT action~\eqref{eq:actionEFT} derived from the most general second-order scalar-tensor theories.
We still have the four independent functions of time, $\alpha_T$, $\alpha_B$, $\alpha_V$, and $\alpha_M$.
By requiring that the speed of gravitational waves is equal to that of light, we set
\begin{align}
    \alpha_T=-\alpha_V=0.
\end{align}
We further assume, for simplicity, that
\begin{align}
    \alpha_M=\frac{1}{H}\frac{\D \ln M}{\D t}=0.
\end{align}
Thus, only a single function $\alpha_B$ is left, which characterizes the kinetic gravity braiding effect that is the characteristic feature of the cubic galileon model~\cite{Nicolis:2008in,Deffayet:2009wt,Deffayet:2010qz,Kobayashi:2010cm}.
In this setup, the coefficient of the nonlinear derivative interaction is characterized by $\alpha_B$.

There is another $\alpha$ function, conventionally denoted by $\alpha_K$, which has dropped from the above effective action due to the quasi-static approximation we made.
In principle, $\alpha_K$ can change the abundance of cosmic voids through the modification of the linear matter power spectrum.
This point will be discussed in more detail in Sec.~\ref{sec:vsf}.

\subsubsection{Background model}

Let us now specify the model of the evolution of the cosmological background.
To focus on studying the impacts of modified gravity on the evolution of underdensities by the use of the EFT of dark energy, we assume that the background evolution is the same as that of the $\Lambda$CDM model.
The Hubble parameter $H(t)$ it therefore obeys
\begin{align}
    \frac{H^2}{H_0^2}=\frac{\Omega_{\mrm r0}}{a^4}+\frac{\omp}{a^3}+1-\omp-\Omega_{\mrm r0},
\end{align}
where $H_0$, $\omp$, and $\Omega_{\mrm r0}$ are the present value of the Hubble parameter, the matter density parameter, and the radiation density parameter, respectively.
Unless otherwise stated, we ignore the radiation-dominated stage in the following analysis.

The Hubble parameter and the scale factor are given explicitly in terms of $t$ by
\begin{align}
    a^3(t)&=\frac{\omp}{1-\omp}\sinh^2\left[\frac{3}{2}(1-\omp)^{1/2}H_0t\right],\\
    H(t)&=H_0(1-\omp)^{1/2}\coth\left[\frac{3}{2}(1-\omp)^{1/2}H_0t\right],
\end{align}
where the normalization of the scale factor is determined so that $a=1$ at the present time.
The time-dependent matter density parameter, defined by
\begin{align}
    \om(t):=\frac{8\pi G\bar{\rho}}{3H^2},
\end{align}
can be expressed conveniently in terms of $\omp$ and the scale factor as
\begin{align}
    \om(t)=\frac{H_0^2\omp}{a^3H^2}=\frac{\omp}{\omp+a^3(1-\omp)}.
\end{align}
Here, we have defined the Newton constant $G$ as $G=1/8\pi M^2$.

In the EFT of dark energy, the time dependence of the $\alpha$ functions needs to be specified.
In this paper, we assume that
\begin{align}
    \alpha_i(t)=\alpha_{i0}\left(\frac{H_0}{H}\right)^{2p}\simeq\alpha_{i0}\left[\frac{1-\om(t)}{1-\omp}\right]^p,
    \quad 
    i=K, B,
    \label{eq:alpha}
\end{align}
where $p$ is some constant and $\alpha_{i0}$ is the present value of $\alpha_i$.
The second equality is valid from the matter-dominated era to the present time.
The model with $p=1$ is most frequently used in the literature.
The model with general $p$ was considered in Ref.~\cite{Traykova:2021hbr}, in which it was reported that this parametrization fits more than $98\%$ of randomly generated shift-symmetric models with constant $G_4$ in the Horndeski action.

\subsubsection{Gravitational-field equations and equation of motion for the shells}
\label{subsec:grav}

Varying the EFT action~\eqref{eq:actionEFT} with respect to $\Phi$, $\Psi$, and $\pi$
and integrating once the resultant equations,
we obtain
\begin{align}
    &\frac{\Phi'}{a^2r}=\frac{\Psi'}{a^2r}
    =-\frac{H\alpha_B}{2}\left(\frac{\pi'}{a^2r}\right)
    +H^2\delta m,\label{eq:veft-1}
\end{align}
and
\begin{align}
    \alpha_B\left(\frac{\pi'}{a^2r}\right)^2-H
    \left(\alpha_*+\frac{\alpha_B^2}{4}\right)\left(\frac{\pi'}{a^2r}\right)
    -\frac{\alpha_B}{2}\left(\frac{\Phi'}{a^2r}\right)=0,
    \label{eq:veft-2}
\end{align}
where 
\begin{align}
    \alpha_*&:=-\frac{3\Omega_{\mrm{m}}}{2}-\frac{\dot H}{H^2}
    +\frac{\dot H\alpha_B}{2H^2}+\frac{\alpha_B}{2}
    \left(1-\frac{\alpha_B}{2}\right)+\frac{\dot\alpha_B}{2H},
    \label{def:alpha-ast}
    \\ 
    \delta m&:=\frac{G\mcal{M}}{H^2R^3}-\frac{\Omega_{\mrm{m}}}{2},
\end{align}
and a prime denotes differentiation with respect to $r$.
Note that for the $\Lambda$CDM background we have $(3/2)\Omega_{\mrm{m}}+\dot H/H^2=0$, and hence the $\alpha_B$-independent part in Eq.~\eqref{def:alpha-ast} in fact vanishes.
With the help of Eq.~\eqref{eq:veft-1} one can eliminate $\Phi'$ from Eq.~\eqref{eq:veft-2} to get a quadratic equation for $\pi'$.
The solution is given by
\begin{align}
    \frac{2\alpha_B}{H}\left(\frac{\pi'}{a^2r}\right)
    =\alpha_*\pm\sqrt{\alpha_*^2+2\alpha_B^2\delta m},
\end{align}
yielding
\begin{align}
    \frac{1}{R}\frac{\partial\Phi}{\partial R}=\frac{\Phi'}{a^2r}=H^2\delta m
    -\frac{H^2}{4}\left(\alpha_*\pm \sqrt{\alpha_*^2+2\alpha_B^2\delta m}\right).
    \label{eq:DPhi-DR}
\end{align}
The upper (lower) sign should be taken for $\alpha_*<0$ ($\alpha_*>0$) so that $\pi',\Phi',\Psi'\to 0$ when the source vanishes, $\delta m\to 0$.
We require that scalar perturbations are free from the ghost instability,
leading to $\alpha_K+(3/2)\alpha_B^2>0$.
This, combined with the condition that the sound speed squared is positive,
\begin{align}
    c_s^2=\frac{2\alpha_*}{\alpha_K+3\alpha_B^2/2}>0,
    \label{eq:def:sound-speed2}
\end{align}
gives $\alpha_*>0$.
We therefore choose the lower sign in the above expressions.
Substituting Eq.~\eqref{eq:DPhi-DR} into Eq.~\eqref{eq:R}, we obtain
\begin{align}
    \frac{\ddot R}{R}=\frac{\ddot a}{a}
    -H^2\delta m
    +\frac{H^2}{4}\left(\alpha_*- \sqrt{\alpha_*^2+2\alpha_B^2\delta m}\right).
    \label{eq:Reft}
\end{align}
For given $H(t)$, $\Omega_{\mrm{m}}(t)$, and $\alpha_B(t)$, this equation can be used to determine the trajectory of a particle $R(t)$ in the EFT of dark energy.
It is worth noting that the expression in the square root, $\alpha_*^2+2\alpha_B^2\delta m$, can be negative in an underdense region, $\delta m<0$, depending on the assumed cosmological background model.
This would result in imaginary $\pi$, and hence
we exclude such cosmological background models.
The same problem has been pointed out already in Refs.~\cite{Winther:2015pta} in the context of cubic Galileon models.

\subsubsection{Linear regime}

If $|\delta|\ll 1 $, we may assume that $|\delta m|\ll 1$ and linearize the governing equations.
Let us denote the linear density contrast by $\delta_L$.
It follows from Eq.~\eqref{eq:DPhi-DR} and $R^{-2}\partial_R\left(R^3H^2\delta m\right)= 4\pi G \bar\rho \delta_L$ that, to linear order in $\delta_L$,
\begin{align}
    \frac{1}{a^2}\nabla^2\Phi = 4\pi G \mu\bar\rho \delta_L,
    \label{eq:gen-poisson}
\end{align}
with
\begin{align}
    \mu := 1+\frac{\alpha_B^2}{4\alpha_*}.
    \label{eq:def:mu}
\end{align}
Since the stability condition amounts to $\alpha_*>0$, the effective gravitational constant for the linear density contrast, $G\mu$, is always enhanced relative to $G$ by the factor $\mu\,(>1)$.
Combining this Poisson equation with the linearized evolution equation for the density contrast,
\begin{align}
    \ddot \delta_L+2H\dot\delta_L=\frac{1}{a^2}\nabla^2\Phi,
\end{align}
we obtain the closed equation for $\delta_L$:
\begin{align}
     \ddot\delta_L+2H\dot\delta_L = 4\pi G \mu\bar\rho \delta_L.
     \label{eq:deltaL}
\end{align}

Since we assume that the time-dependence of $\alpha_B$ is given by Eq.~\eqref{eq:alpha},
we have $\alpha_B\to 0$ in the deep matter-dominated era, $\Omega_\mrm{m}\to 1$.
Therefore, the standard equation in the Einstein-de Sitter universe is recovered in the deep matter-dominated era,
\begin{align}
\ddot\delta_L+\frac{4}{3t}\dot\delta_L=\frac{2}{3t^2}\delta_L,
\label{eq:linear}
\end{align}
giving the standard growing solution in that era,
\begin{align}
\delta_L\propto a.
\end{align}

\section{Evolution of voids}\label{sec:void}

\subsection{Initial profile of the underdense region and shell-crossing}\label{subsec:shellcross}

Let us consider an isolated spherical void whose initial density profile at $t=t_i$ is given by
\begin{align}
  \delta(t_i,R_i)=\begin{cases}
    -\delta_0&\quad\mrm{for}\quad R_i<R_0-\Delta R\\
    F(R_i)&\quad\mrm{for}\quad R_0-\Delta R\leq R_i\leq R_0+\Delta R\\
    0&\quad\mrm{for}\quad R_i>R_0+\Delta R,
  \end{cases}
\end{align} 
where $\delta_0\,(>0)$ is the amplitude of the initial density contrast in the underdense region.
In $R_0-\Delta R\le R\le R_0+\Delta R$, we insert the (thin) region interpolating the inner underdense and outer regions.
The function $F(R)$ used for the interpolation is given by
\begin{align}
  F(R)=\eta_3R^3+\eta_2R^2+\eta_1R+\eta_0,
\end{align}
where the coefficients $\eta_0,\dots,\eta_3$ are determined by the conditions for smooth matching:
\begin{align}
  F(R_0-\Delta R)=-\delta_0,\quad F(R_0+\Delta R)=0,\quad F'(R_0-\Delta R)=0,\quad F'(R_0+\Delta R)=0.
\end{align}
To quantify the width of this intermediate region, we define the sharpness parameter $s$ as
\begin{align}
  s:=\frac{R_0}{\Delta R}.
\end{align}

\begin{figure}[tb]
  \centering
  \includegraphics[keepaspectratio,scale=0.7]{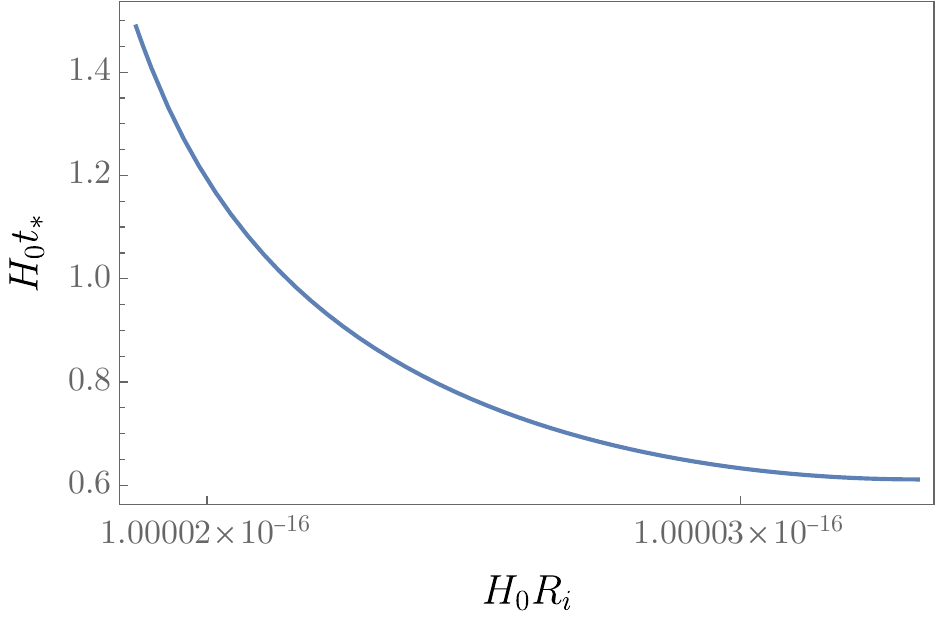}
  \caption{Shell-crossing time of each shell $t_*$ versus the initial physical radius $R_i$ in the $\Lambda$CDM model in general relativity, assuming $\omp=0.3153$ and the initial profile characterized by $R_0=10^{-16}H_0^{-1}$, $s=3\times10^4$, and $\delta_0=3\times10^{-7}$.}
  \label{fig:sctime}
\end{figure}

To determine the moment of void formation, we first need to specify the condition that signals the formation of a void.
A common criterion is that a void forms when shell-crossing occurs, i.e., two neighboring shells intersect.
The moment of shell-crossing is determined from the condition~\cite{Moretti:2025gbp}
\begin{align}
  \lim_{\varepsilon\to0^+}\frac{R(t,R_i+\varepsilon)-R(t,R_i)}{\varepsilon}=0,
  \label{def:shellcross}
\end{align}
where $R_i$ is the initial physical radius of the inner one of the two shells and $R(t,R_i)$ is the physical radius of that shell at the time $t$.
In general, the time $t=t_*$ satisfying Eq.~\eqref{def:shellcross} depends on $R_i$: $t_*(R_i)$.
Following Ref.~\cite{Moretti:2025gbp}, we adopt the definition that the void forms at the moment $t=t_{\mrm{sc}}$ when Eq.~\eqref{def:shellcross} is satisfied for the outermost shells (i.e., $t_{\mrm{sc}}:=t_*(R_0+\Delta R)$).
As shown in Fig.~\ref{fig:sctime}, for $s\gg 1$, the two outermost shells intersect first for our choice of the initial conditions specified below.
With this definition of $t_{\mrm{sc}}$, the known analytical result in the Einstein-de Sitter universe, $\delta_{\mrm{sc}}\simeq-2.718$, is reproduced when $s\gg 1$ is taken, which supports adopting the above definition in a more general setup~\cite{Moretti:2025gbp}.

\subsection{Evolution of spherical underdensities and overdensities}\label{subsec:evol}

\subsubsection{Underdensity}

\begin{figure}[tb]
  \begin{minipage}[b]{0.45\linewidth}
    \centering
    \includegraphics[keepaspectratio, scale=0.45]{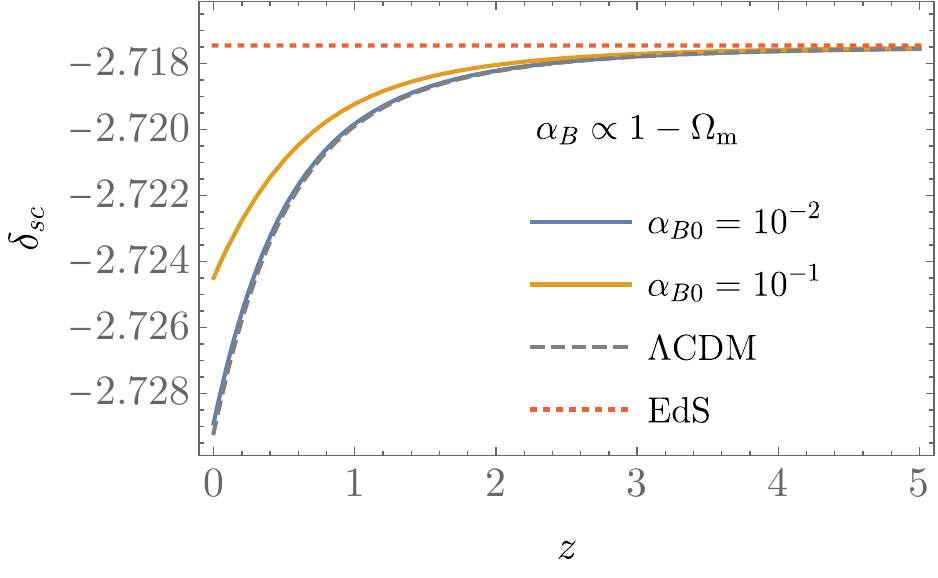}
  \end{minipage} 
  \begin{minipage}[b]{0.45\linewidth}
    \centering
    \includegraphics[keepaspectratio, scale=0.45]{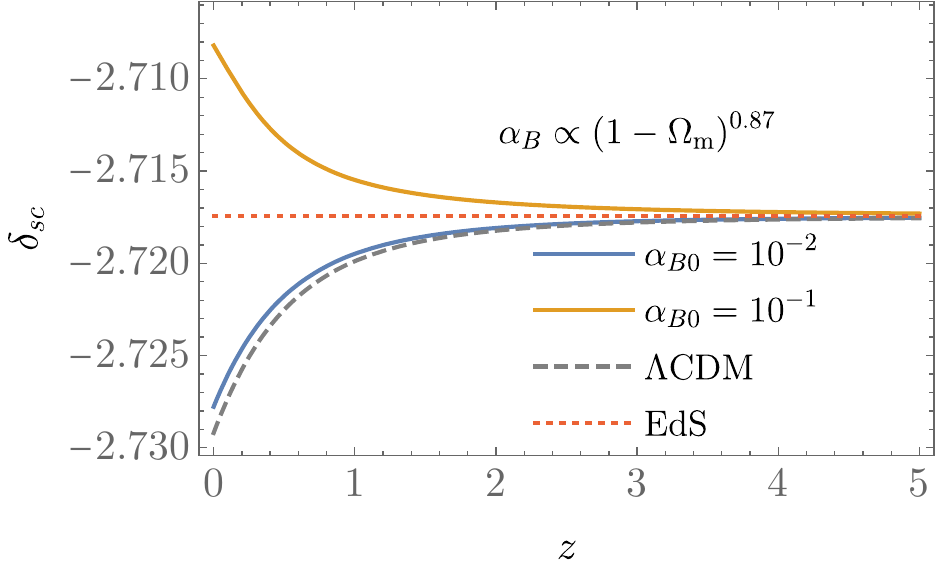}
  \end{minipage} \\
  \caption{Linear density contrast at the shell-crossing time in the EFT of dark energy with $p=1$ and $p=0.87$. The parameter is given by $\alpha_{B0}=10^{-2}$ (solid blue line) and $\alpha_{B0}=10^{-1}$ (solid orange line). The dashed and dotted lines represent the results in the $\Lambda$CDM model and Einstein-de Sitter model, respectively.}
 \label{fig:delta_sc}
\end{figure}

\begin{figure}[tb]
  \begin{minipage}[b]{0.45\linewidth}
    \centering
    \includegraphics[keepaspectratio, scale=0.45]{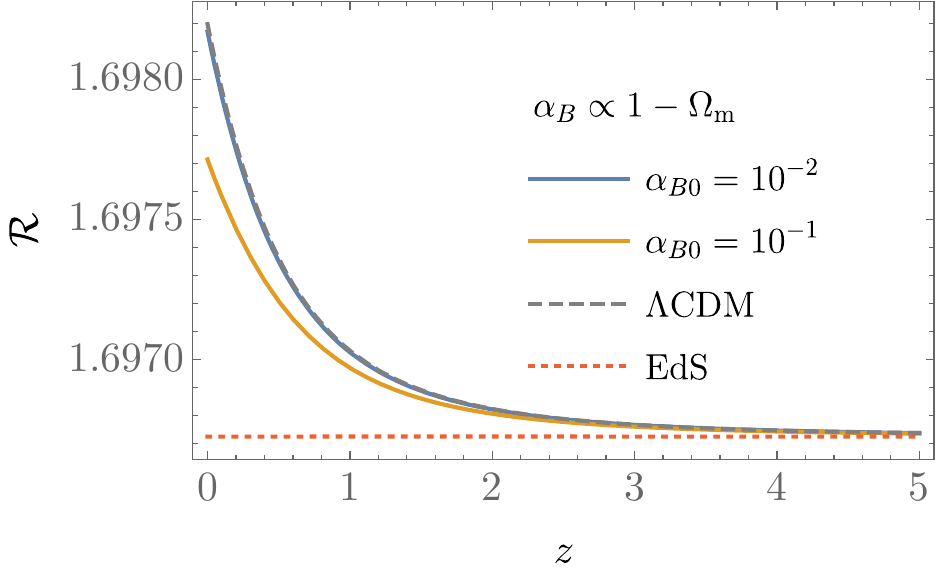}
  \end{minipage} 
  \begin{minipage}[b]{0.45\linewidth}
    \centering
    \includegraphics[keepaspectratio, scale=0.45]{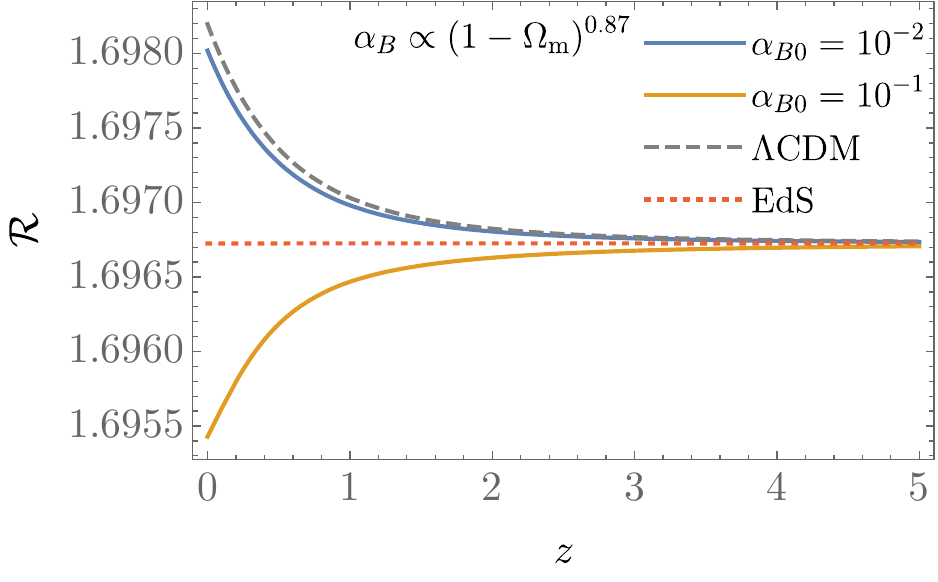}
  \end{minipage} \\
  \caption{Expansion factor in the EFT of dark energy with $p=1$ and $p=0.87$. The parameter is given by $\alpha_{B0}=10^{-2}$ (solid blue line) and $\alpha_{B0}=10^{-1}$ (solid orange line). The dashed and dotted lines represent the results in the $\Lambda$CDM and Einstein-de Sitter models, respectively.}
 \label{fig:expansion}
\end{figure}

Now we are ready to compute the nonlinear evolution of the underdense region and estimate the time of void formation.
To this end, we solve Eq.~\eqref{eq:Reft} with appropriate initial conditions and determine the shell-crossing time of the outermost shell.
We set the initial conditions for $R$ as
\begin{align}
    R(t_i)=R_{i,\mrm{out}},\quad\dot{R}(t_i)=H(t_i)R_{i,\mrm{out}}-\frac{1}{3}H(t_i)R_{i,\mrm{out}}\Delta(t_i,r_{i,\mrm{out}}),
    \label{eq:Rini}
\end{align}
at $t=t_i$ corresponding to $a(t_i)=10^{-7}$, where $r_{i,\mrm{out}}$ and $R_{i,\mrm{out}}$ are the comoving and physical coordinates of the outermost shell, respectively.
The initial velocity of the shell is given by the sum of the Hubble flow and the peculiar velocity, and $\Delta(t,R)$ is the average density contrast.
The derivation of the above expression for the peculiar velocity and the precise definition of $\Delta(t,r)$ are found in Appendix~\ref{app:velocity}.
The parameters for specifying the initial profile of $\delta(t,R)$ are set as
\begin{align}
    R_0=10^{-16}H_0^{-1},\quad s=3 \times 10^4.
\end{align}
Solving the evolution equation~\eqref{eq:Reft} for $R$ with the initial conditions~\eqref{eq:Rini}, we obtain the shell-crossing time $t_{\mrm{sc}}$ for different assumptions on the time dependence of $\alpha_B$ (i.e., the models with $p=1$ and $p=0.87$).

Having obtained $t_{\mrm{sc}}$, we compute the linearly extrapolated critical density contrast for the void formation, $\delta_{\mrm{sc}}$, defined as
\begin{align}
    \delta_{\mrm{sc}}:=\delta_L(t_{\mrm{sc}}).
\end{align}
This can be obtained by solving Eq.~\eqref{eq:deltaL} with the initial conditions
\begin{align}
    \delta_L(t_i)=A_{Li}a(t_i),\quad\dot{\delta}_L(t_i)=H(t_i)\delta(t_i),
    \label{eq:dLini}
\end{align}
where $A_{Li}=-\delta_0/a(t_i)$.
The resultant $\delta_{\mrm{sc}}$ is shown in Fig.~\ref{fig:delta_sc} as a function of the redshift $z=a^{-1}(t_{\mrm{sc}})-1$ at shell-crossing.
In both $p=1$ and $p=0.87$ models, it can be seen that the critical density contrast increases as $\alpha_{B0}$ increases when compared at the same redshift.
This can be understood as follows: as $\alpha_{B0}$ increases, gravity becomes stronger and the shell-crossing occurs earlier, which tends to make $|\delta_{\mrm{sc}}|$ smaller.
In contrast, the linear growth of $\delta_L$ is enhanced with increasing $\alpha_{B0}$, which acts to make $|\delta_{\mrm{sc}}|$ larger.
The final value of $\delta_{\mrm{sc}}$ is determined by the competition between these two effects.
It is found that the two effects are comparable and nearly cancel each other out.
As a result, the modification of $\delta_{\mathrm{sc}}$ is one order of magnitude smaller than the order of $\alpha_{B0}$.
Since we have $\Omega_{\mathrm{m}}\to 1$ and $\alpha_{B}\to 0$ in the deep matter-dominated era, $\delta_{\mrm{sc}}$ approaches the Einstein-de Sitter value at early times.

To compute the void size function, we also need to evaluate the expansion factor $\mathcal{R}$ defined by
\begin{align}
\mcal R(t) := \frac{R(t_{\mrm{sc}})/a(t_{\mrm{sc}})}{R(t_i)/a(t_i)},
\end{align}
which quantifies how much the comoving radius of a shell expands by the shell-crossing time.
Figure~\ref{fig:expansion} shows the numerical results for the expansion factor $\mcal R$ as a function of redshift $z$, for the models with $p=1$ and $p=0.87$.
We find that the expansion factor decreases as $\alpha_{B0}$ increases.
This is because a larger value of $\alpha_{B0}$ strengthens gravity, thereby slowing down the expansion of each shell.
Consequently, the comoving radius of the shell at the shell-crossing time becomes smaller.

One might be interested in how $\delta_{\mathrm{sc}}$ is modified in models with smaller values of $p$.
However, we find that it is not possible to compute the shell evolution for $p\lesssim0.87$ with the fiducial value $\alpha_{B0}=0.1$, because the term inside the square root in Eq.~\eqref{eq:Reft} becomes negative at some moment, causing the physical radius of the shell to become imaginary before the shell-crossing occurs for any initial amplitude $A_{Li}$ of $\delta$.
Note that this occurs because we are considering an underdense region ($\delta m<0$).

We only investigate the case with $\alpha_{B0}>0$ in this paper, because gradient instabilities show up in our setup for $\alpha_{B0}<0$.
This can be seen from the expression for the sound speed squared in the radiation-dominated era,
\begin{align}
    c_s^2=\frac{\widetilde\alpha_*}{\alpha_K+3\alpha_B^2/2},
\end{align}
where
\begin{align}
    \widetilde\alpha_* := -2\Omega_{\mrm r}-\frac{\dot{H}}{H^2}+\frac{\dot{H}}{H^2}\alpha_B+\frac{\alpha_B}{2}\left(1-\frac{\alpha_B}{2}\right)+\frac{\dot{\alpha_B}}{2H},
\end{align}
and $\Omega_{\mrm r}(t)$ is the time-dependent radiation density parameter defined as
\begin{align}
    \Omega_{\mrm r}(t) := \frac{8\pi G\bar\rho_{\mrm r}}{3H^2},
\end{align}
with $\bar\rho_{\mrm r}$ denoting the mean radiation energy density.
Using the background equations together with Eq.~\eqref{eq:alpha}, $\widetilde\alpha_*$ reduces, in the deep radiation-dominated era, $\Omega_{\mrm r}\rightarrow1$, to
\begin{align}
    \widetilde\alpha_* = 2\left(p-\frac{1}{4}\right)\alpha_B,
\end{align}
to linear order in $\alpha_B$, which is much smaller than unity in this regime.
To avoid gradient instabilities, we require $\widetilde\alpha_*>0$,
which implies either (i) $p>1/4$ and $\alpha_B>0$ or (ii) $p<1/4$ and $\alpha_B<0$.
We do not consider case (ii) because an imaginary scalar field occurs as far as we have examined.

\subsubsection{Overdensity}
\label{sec:overdensity}

\begin{figure}[tb]
  \begin{minipage}[b]{0.45\linewidth}
    \centering
    \includegraphics[keepaspectratio, scale=0.45]{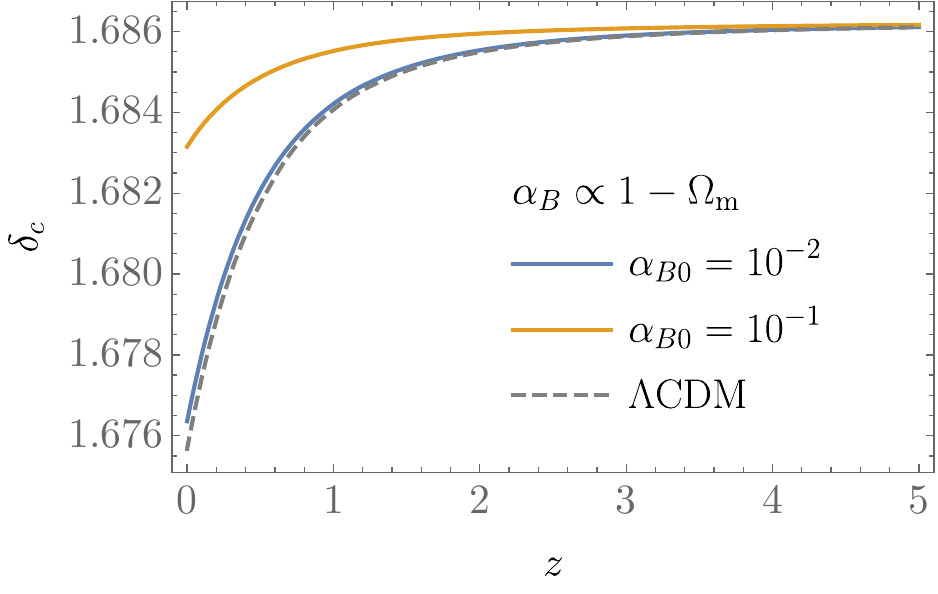}
  \end{minipage} 
  \begin{minipage}[b]{0.45\linewidth}
    \centering
    \includegraphics[keepaspectratio, scale=0.45]{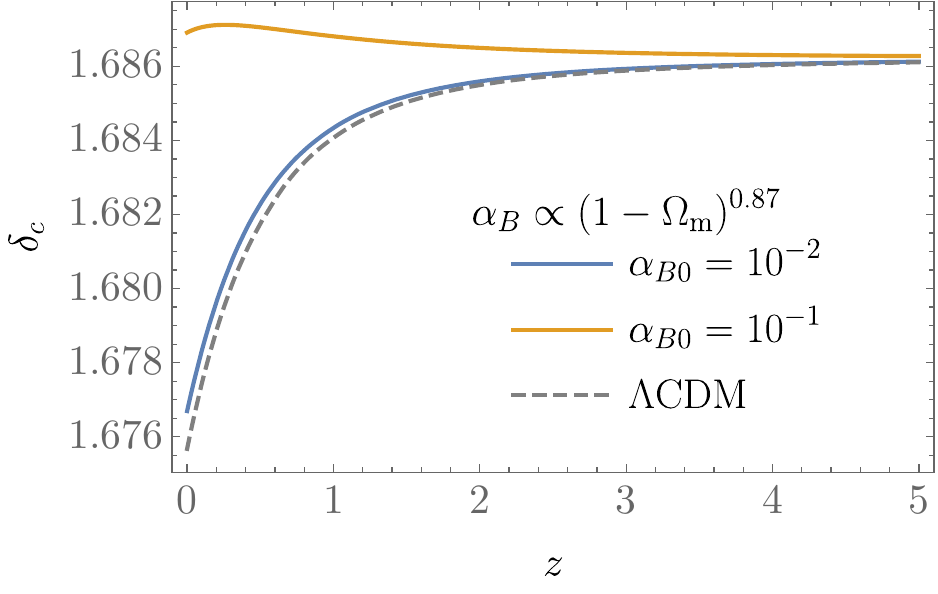}
  \end{minipage} \\
  \caption{Linear density contrast at the collapse time in the EFT of dark energy with $p=1$ and $p=0.87$. The parameter is given by $\alpha_{B0}=10^{-2}$ (solid blue line) and $\alpha_{B0}=10^{-1}$ (solid orange line). The dashed and dotted lines represent the results in the $\Lambda$CDM and Einstein-de Sitter models, respectively.}
 \label{fig:delta_c}
\end{figure}

We need to calculate the critical density contrast $\delta_{\mathrm{c}}$ for the halo formation to account for the void-in-cloud problem, as reviewed in Appendix.~\ref{app:excursion}.

For the nonlinear evolution of a top-hat overdensity, we solve Eq.~\eqref{eq:delta} with the initial conditions $\delta(t_i)=A_ia(t_i)$ and $\dot{\delta}(t_i)=H(t_i)\delta(t_i)$,
where $A_i$ is a constant.
These initial conditions are justified because the linear evolution of $\delta$ is governed by the standard equation in the deep matter-dominated era.
We suppose that the overdense region collapses when $\delta$ reaches $10^7$, and evaluate the collapse time $t_{\mrm{col}}$ using the condition $\delta(t_{\mrm{col}})=10^7$.
Then, we calculate the critical density contrast $\delta_{\mathrm{c}}$ from the linear evolution, $\delta_{\mathrm{c}}=\delta_L(t_{\mrm{col}})$.
The resultant critical density contrast $\delta_{\mathrm{c}}$ is shown in Fig.~\ref{fig:delta_c} as a function of the redshift $z=a^{-1}(t_{\mrm{col}})-1$ for the background model with $p=1$ and $p=0.87$.

\section{Implications for the void size function}\label{sec:vsf}

To compare theoretical predictions with observational data of cosmic voids, it is crucial to study the VSF.
To this end, we adopt the VSF proposed in Ref.~\cite{Sheth:2003py},
\begin{align}
\frac{\mathrm{d}n(r,z)}{\mathrm{d}\ln r}
=\left.
\frac{f(\sigma(r_L,z),\delta_{\mathrm{c}}(z),\delta_{\mrm{sc}}(z))}{V(r_L)}
\frac{\mathrm{d}\ln\sigma^{-1}}{\mathrm{d}\ln r_L}\right|_{r_L=r/\mcal R(z)},
\end{align}
where $n(r,z)$ is the number density of voids with comoving radius $r$ at redshift $z$, $V(r_L)=4\pi r_L^3/3$, $\mcal R$ is the expansion factor introduced in Sec.~\ref{subsec:evol}, and $\sigma^2$ is the variance of the linear density field smoothed on scale $r_L$:
\begin{align}
  \sigma^2(r_L,z)=\frac{1}{2\pi^2}\int\D k k^2W^2(k,r_L)P(k,z),
\end{align}
with the Fourier-space top-hat window,
\begin{align}
    W(k,r_L):=\begin{cases}
        &1\qquad k\leq r_L^{-1}\\
        &0\qquad k> r_L^{-1}
    \end{cases},
\end{align}
and the linear matter power spectrum at redshift $z$, $P(k,z)$.
The explicit form of the function $f(\sigma,\delta_{\mathrm{c}},\delta_{\mrm{sc}})$ will be given below.
The quantities $\delta_{\mrm{sc}}$, $\mcal{R}$, and $\delta_{\mathrm{c}}$ are computed in the way described in Sec.~\ref {subsec:evol}.
The matter power spectrum is computed using a modified version of the publicly available Einstein--Boltzmann solver \texttt{hi{\_}class}~\cite{Zumalacarregui:2016pph}.
When computing the power spectrum, we adopt the fiducial value $\alpha_{K0}=1$, unless stated otherwise.
We are thus ready to calculate the VSF.

\begin{figure}[tb]
  \begin{minipage}[b]{0.45\linewidth}
    \centering
    \includegraphics[keepaspectratio, scale=0.45]{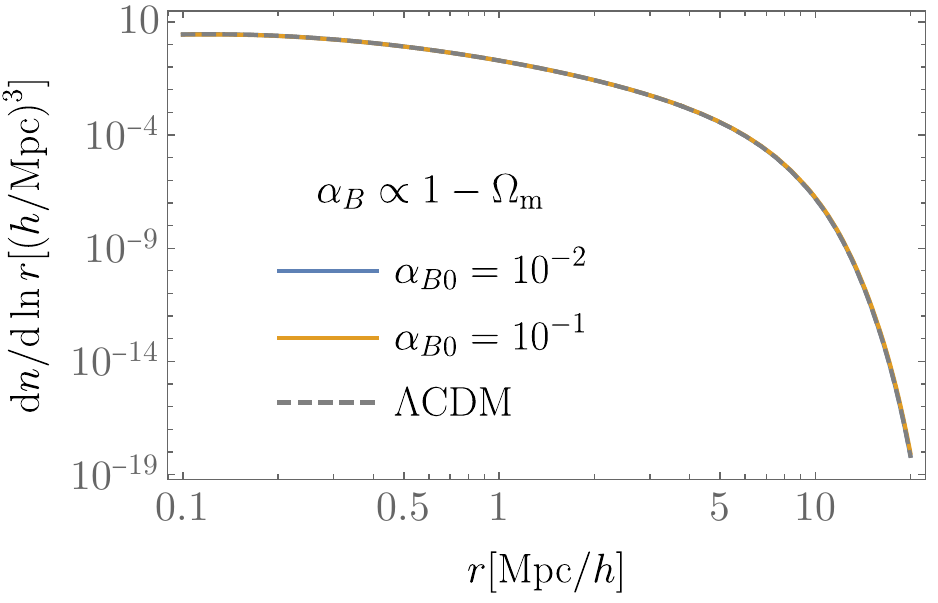}
  \end{minipage} 
  \begin{minipage}[b]{0.45\linewidth}
    \centering
    \includegraphics[keepaspectratio, scale=0.45]{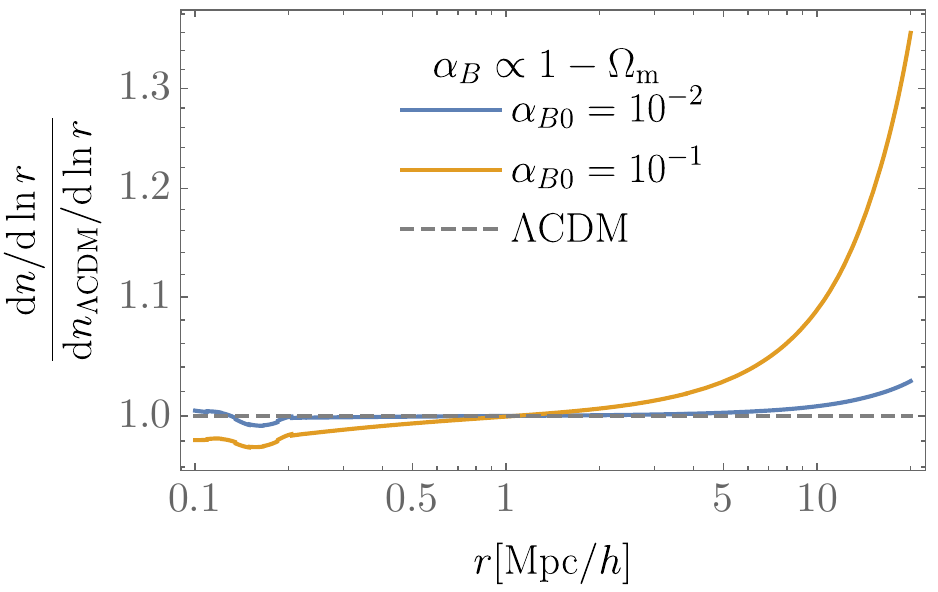}
  \end{minipage} \\
  \caption{Void size function (at $z=0$) in the EFT of dark energy (left) and the relative differences from that in the $\Lambda$CDM model (right),
  with the parameter being given by $\alpha_{B0}=10^{-2}$ (solid blue line) and $\alpha_{B0}=10^{-1}$ (solid orange line).
  The time-dependence of the EFT parameters is described by the model with $p=1$.
  The dashed lines represent the result in the $\Lambda$CDM model.}
 \label{fig:vsf1}
\end{figure}

\begin{figure}[tb]
  \centering
  \includegraphics[keepaspectratio,scale=0.7]{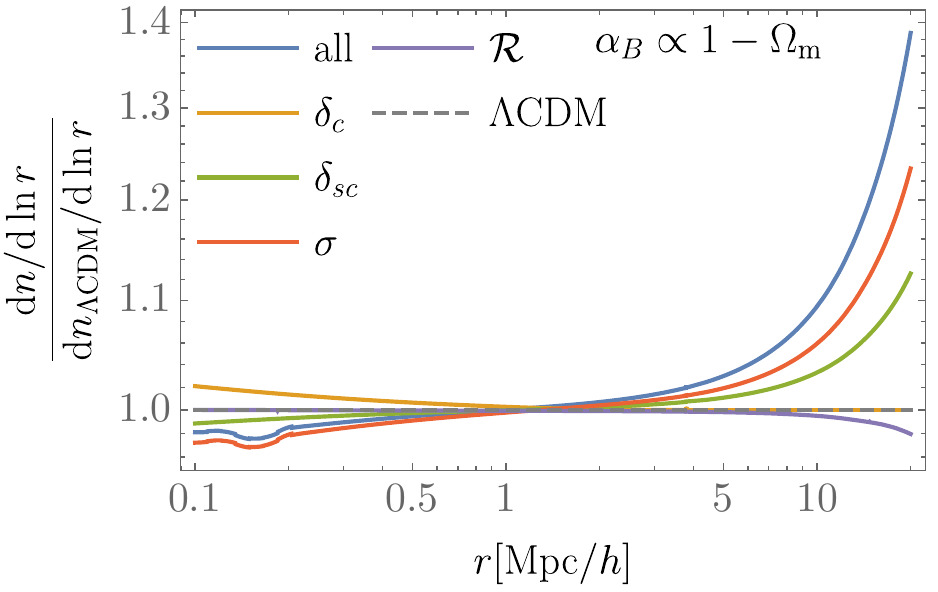}
  \caption{Relative differences between the VSF in the EFT of dark energy and that in the $\Lambda$CDM model (at $z=0$). To isolate the individual effect of the modification of gravity, we turn on each of the components $(\sigma,\delta_{\mathrm{c}},\delta_{\mrm{sc}},\mcal{R})$ one at a time
  for the model with $p=1$ and $\alpha_{B0}=0.1$.}
  \label{fig:vsf_compare}
\end{figure}

\begin{figure}[tb]
  \begin{minipage}[b]{0.45\linewidth}
    \centering
    \includegraphics[keepaspectratio, scale=0.45]{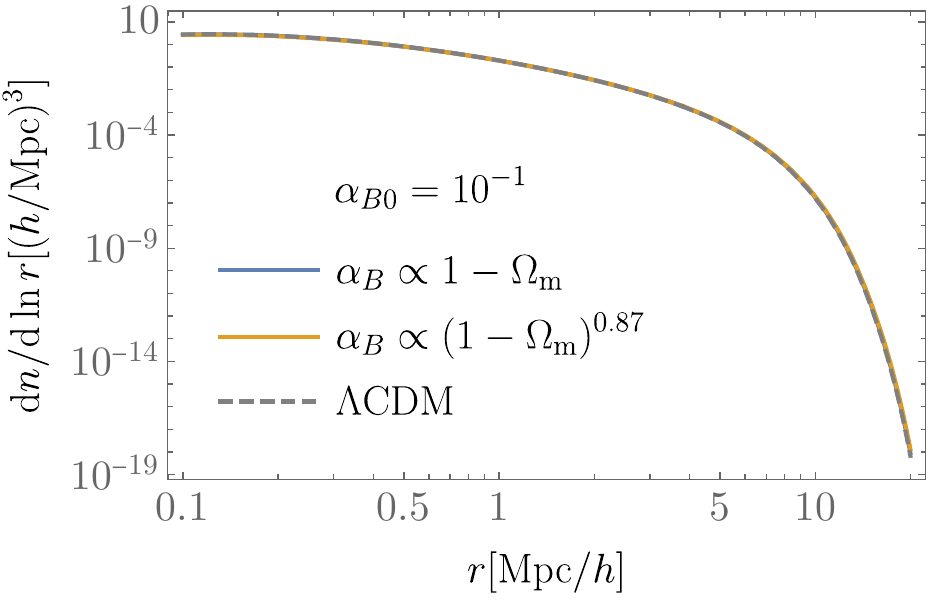}
  \end{minipage} 
  \begin{minipage}[b]{0.45\linewidth}
    \centering
    \includegraphics[keepaspectratio, scale=0.45]{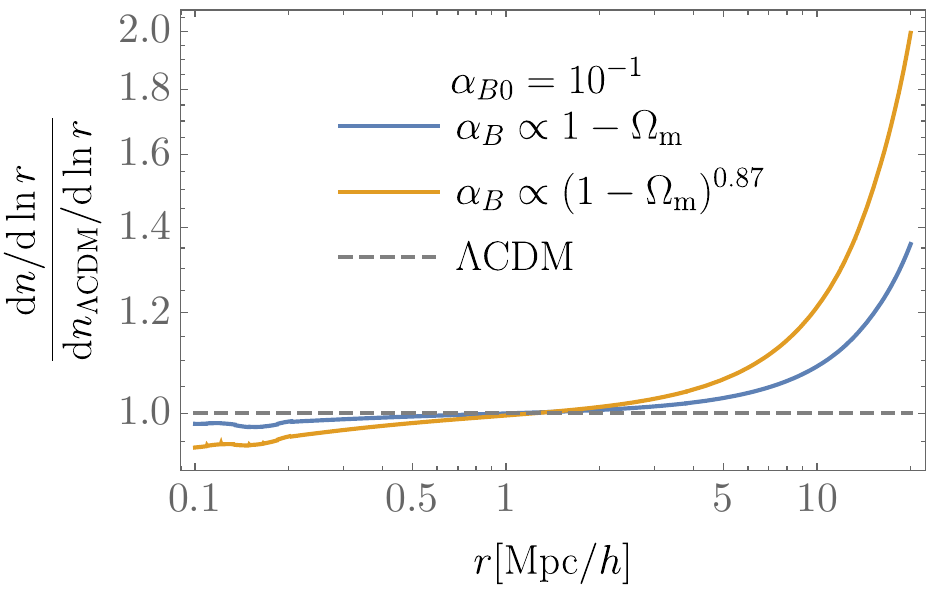}
  \end{minipage} \\
  \caption{Void size function for $p=1$ (solid blue line) and $p=0.87$ (solid orange line) at $z=0$ (left) and the relative differences from that in the $\Lambda$CDM model (right).
  The EFT parameter is fixed as $\alpha_{B0}=0.1$.
  The dashed lines represent the results in the $\Lambda$CDM model.}
 \label{fig:vsf2}
\end{figure}

The explicit form of function $f(\sigma,\delta_{\mathrm{c}},\delta_{\mrm{sc}})$ is given by~\cite{Sheth:2003py,Jennings:2013nsa}
\begin{align}
  f(\sigma,\delta_{\mathrm{c}},\delta_{\mrm{sc}})=2\sum^{\infty}_{j=1}
  e^{-(j\pi x)^2/2}j\pi x^2\sin(j\pi\mcal{D}),
\end{align}
where
\begin{align}
  \mcal{D}:=\frac{|\delta_{\mrm{sc}}|}{\delta_{\mathrm{c}}+|\delta_{\mrm{sc}}|},\qquad x:=\frac{\mcal{D}}{|\delta_{\mrm{sc}}|}\sigma.
\end{align}
It is not easy to compute the above infinite sum.
However, in Ref.~\cite{Jennings:2013nsa}, the authors showed that if $\mcal{D}$ is smaller than $3/4$, this function can be approximated as
\begin{align}
  f(\sigma,\delta_{\mathrm{c}},\delta_{\mrm{sc}})\simeq\begin{cases}
    \displaystyle{\sqrt{\frac{2}{\pi}}\frac{|\delta_{\mrm{sc}}|}{\sigma}e^{-\delta_{\mrm{sc}}^2/2\sigma^2}}
    &\quad\mathrm{for}\quad x\leq0.276\\
    \displaystyle{2\sum^{4}_{j=1}e^{-(j\pi x)^2/2}j\pi x^2\sin(j\pi\mcal{D})}
    &\quad\mathrm{for}\quad x>0.276
  \end{cases}.
\end{align}
We use this approximation to compute the VSF.

First, we investigate the impact of $\alpha_{B0}$ on the VSF in the background model with $p=1$.
Figure~\ref{fig:vsf1} shows the VSF and its relative deviation from that of the $\Lambda$CDM model evaluated at the present time.
We find that the VSF exhibits a scale-dependent modification in this model.
On small scales, the VSF for $\alpha_{B0}=0.1$ is suppressed by $\sim 5\%$ compared to that in the $\Lambda$CDM model at $r\sim 0.1\,\mathrm{Mpc}/h$.
Conversely, on large scales, the VSF is enhanced by $\sim 40\%$ at $r\sim 20\,\mathrm{Mpc}/h$ for the same parameters.

Next, to identify which component of $(\sigma,\delta_{\mathrm{c}},\delta_{\mrm{sc}},\mcal{R})$ is responsible for the changes in the VSF, we compute the VSF by using one of the components calculated in the theory with $\alpha_{B0}=10^{-1}$ while using the rest of the components calculated in the $\Lambda$CDM model.
The results are presented in Fig.~\ref{fig:vsf_compare}.
From this figure, two features can be noted.
On small scales, the change in the VSF comes mainly from the modification in the linear matter power spectrum, since the changes that come from $\delta_{\mrm c}$ and $\delta_{\mrm{sc}}$ compensate.
In contrast, on large scales, the modifications in $\sigma$ and $\delta_{\mrm{sc}}$ together account for the change in the VSF, though the modified linear matter power spectrum still has the largest impact.
On all scales, the modifications in $\mcal R$ and $\delta_c$ have negligible impacts on the VSF.

Let us then discuss the impact of the parameter $p$.
We compare the VSFs for $p=1$ and $p=0.87$, while fixing $\alpha_{B0}=0.1$.
Figure~\ref{fig:vsf2} shows that the change in the VSF is enhanced as $p$ decreases.
This behavior arises because the modification parametrized by $\alpha_B$ comes into play earlier for $p=0.87$ than for $p=1$.
For $p>1$, the VSF becomes closer to that of the $\Lambda$CDM model, since the effect of $\alpha_B$ is more strongly suppressed at early times for larger $p$.

\begin{figure}[tb]
  \begin{minipage}[b]{0.45\linewidth}
    \centering
    \includegraphics[keepaspectratio, scale=0.45]{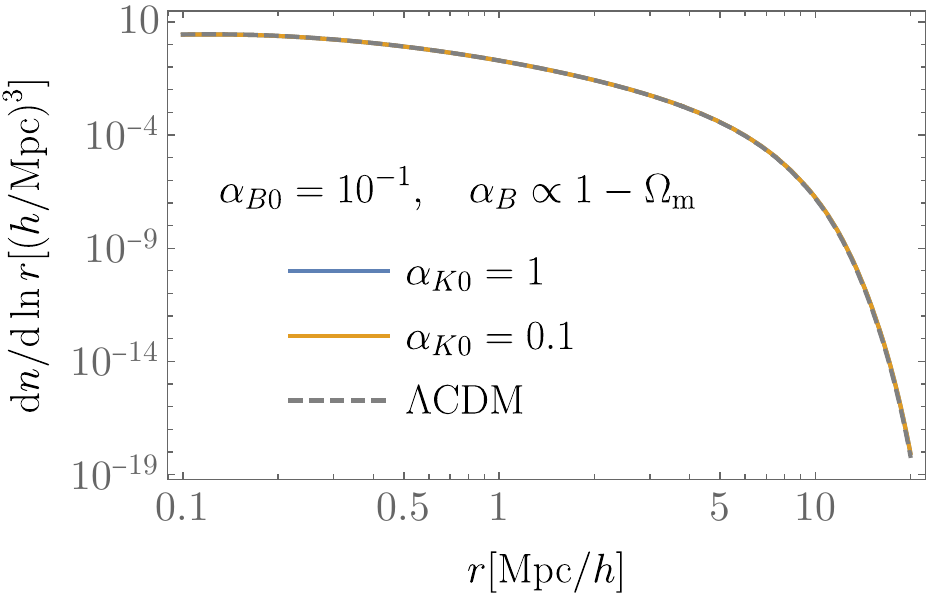}
  \end{minipage} 
  \begin{minipage}[b]{0.45\linewidth}
    \centering
    \includegraphics[keepaspectratio, scale=0.45]{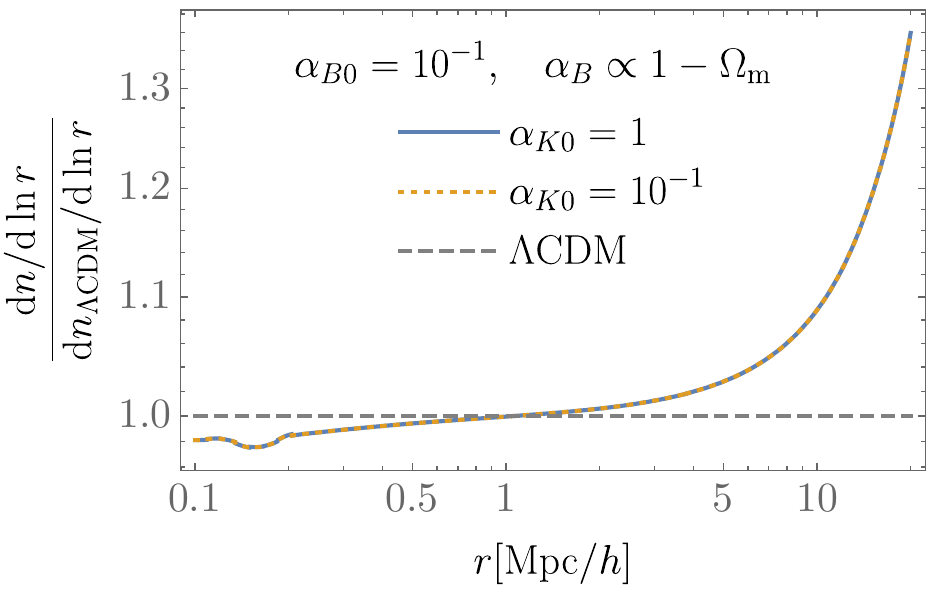}
  \end{minipage} \\
  \caption{Void size function (left) and the relative differences from that in the $\Lambda$CDM model (right). In these plots, the impacts of changing the EFT parameter $\alpha_K$ are investigated.
  The parameter is given by $\alpha_{K0}=1$ (solid blue line) and $\alpha_{K0}=0.2$ (dotted orange line).
  The dashed lines represent the results in the $\Lambda$CDM model.}
 \label{fig:vsf3}
\end{figure}

Finally, let us comment on the impact of $\alpha_{K0}$ on the VSF.
In principle, varying $\alpha_{K0}$ affects the sound speed $c_s$ of the scalar perturbations through Eq.~\eqref{eq:def:sound-speed2}, thereby modifying the VSF.
However, our numerical results shown in Fig.~\ref{fig:vsf3} imply that the VSF is insensitive to $\alpha_{K0}$.
This is because varying $\alpha_{K0}$ leaves the variance almost unchanged.

\section{Conclusions}\label{sec:conclusions}

In this paper, we have studied the evolution of isolated spherical underdense regions and the abundance of cosmic voids within the framework of the effective field theory (EFT) of dark energy.
Using the EFT of dark energy allows us to investigate these phenomena model-independently without specifying the explicit form of the functions in the Horndeski action for dark energy/modified gravity.
Instead, the background evolution was assumed to be the same as that of the conventional $\Lambda$CDM model, and a particular functional form was employed for the time-dependent EFT coefficients.
We have focused on the subset of the Horndeski family in which the speed of the gravitational waves is equal to that of light and the effective Planck mass is time-independent.
Such theories can be characterized by the two EFT coefficients, commonly denoted $\alpha_K$ and $\alpha_B$ in the literature, and we assumed that their time-dependence is given in terms of the matter density parameter by
$\alpha_i\propto [1-\Omega_{\mrm{m}}(t)]^p$, with $p$ being a constant.
We thus have three parameters, $\alpha_{K0}$, $\alpha_{B0}$, and $p$, where $\alpha_{K0}$ and $\alpha_{B0}$ are the present values of $\alpha_K$ and $\alpha_B$, respectively.
We have investigated the impacts of these parameters on the void formation.

We have tracked the evolution of mass shells moving in a modified gravitational potential, and determined the time of shell-crossing $t=t_{\mrm{sc}}$, i.e., the moment when the two neighboring shells at the outer boundary of the void intersect.
Only the EFT parameter $\alpha_B$ comes into play in the dynamics of mass shells.
Defining a cosmic void to form at the moment of shell-crossing, we have computed the linearly extrapolated density contrast at $t=t_{\mrm{sc}}$, $\delta_{\mrm{sc}}$, as well as the expansion factor $\mcal R$ measuring how much the comoving radius of the outermost shell has expanded by the time of shell-crossing.
Our results demonstrate the following qualitative trends.
For larger $\alpha_{B0}\,(>0)$, $|\delta_{\mrm{sc}}|$ is more suppressed.
This results from a slight imbalance between two competing effects: stronger gravity due to nonvanishing $\alpha_B$ makes shell-crossing to occur earlier, while enhancing the linear growth of density perturbations.
For larger $\alpha_{B0}\,(>0)$, $\mcal{R}$ is more suppressed, as stronger gravity slows down the expansion of each shell.
These quantities are more suppressed for smaller $p$, because $\alpha_B$ is larger in the early time (for fixed $\alpha_{B0}$).

In addition to the above qualitative trends, we have noted that there are theoretical bounds on the parameters coming from the requirement that the scalar field must take a real value during the course of the void formation.
For example, this requirement gives $p\gtrsim 0.87$ for $\alpha_{B0}=0.1$.

Having obtained $\delta_{\mrm{sc}}$, we have computed the void size function (VSF) using the Sheth--van de Weygaert model and investigated the impacts of $\alpha_{B0}$, $\alpha_{K0}$, and $p$ on it.
We have found that the VSF exhibits a scale-dependent modification when $\alpha_{B}$ is switched on.
By examining the individual modifications separately, we have shown that most of the scale dependence in the modified VSF originates from the modification of the linear matter power spectrum.
We also studied the dependence of the VSF on the parameter $p$, and found that the modification of the VSF is enhanced for smaller $p$.
 Although only $\alpha_B$ takes part in the dynamics of mass shells and the calculation of $\delta_{\mrm{sc}}$, the linear matter power spectrum, and hence the VSF, could depend on $\alpha_K$ as well.
 However, it is found that the VSF is insensitive to $\alpha_{K0}$, because the variance turns out to be almost independent of $\alpha_{K0}$.

In this paper, we have focused on the impacts of the EFT parameters characterizing Horndeski gravity.
It would be interesting to extend the present analysis to more general theories of gravity, such as degenerate higher-order scalar-tensor theories~\cite{Langlois:2015cwa}, by incorporating additional EFT parameters.
It is fair to say that many aspects of actual complex process of the void formation are ignored in the simple spherical model, and hence our study is a first step toward understanding in more detail the formation of voids in the EFT of dark energy and modified gravity.
To this end, it is important to perform $N$-body simulations.
Nevertheless, we believe that the simplified treatment with spherical symmetry presented in this paper helps us to understand, at least qualitatively, how the void formation is modified in the EFT of dark energy.

\acknowledgments
We thank Shun Arai for interesting discussions.
The work of TK was supported by
JSPS KAKENHI Grant No.~JP25K07308 and
MEXT-JSPS Grant-in-Aid for Transformative Research Areas (A) ``Extreme Universe'',
No.~JP21H05182 and No.~JP21H05189.

\appendix 
\section{Relation between the Horndeski theory and the EFT of dark energy}
\label{app:EFT}

The action for the Horndeski theory~\cite{Horndeski:1974wa} up to quadratic terms of the second derivatives of the scalar field $\phi$ is given by~\cite{Deffayet:2011gz,Kobayashi:2011nu}
\begin{align}
    S=\int \D^4x\sqrt{-g}\left\{G_2(\phi,X)-G_3(\phi,X)\Box\phi+G_4(\phi,X)^{(4)}R+G_{4X}(\phi,X)\left[(\Box\phi)^2-\phi_{\mu\nu}\phi^{\mu\nu}\right]\right\},
\end{align}
where $\phi_{\mu}:=\cd_\mu\phi$, $\phi_{\mu\nu}:=\cd_\mu\cd_\nu\phi$, $X:=-\phi_{\mu}\phi^{\mu}/2$, and $G_2$, $G_3$, and $G_4$ are arbitrary functions of $\phi$ and $X$. We write $f_X=\pd f/\pd X$.

The $\alpha$-basis formulation of the EFT of dark energy is particularly useful for comparing theoretical predictions with observational data.
Following Refs.~\cite{Bellini:2014fua,Cusin:2017mzw} (see also~\cite{Gleyzes:2014rba,Langlois:2017mxy,Dima:2017pwp}), one can relate the above Horndeski action to the EFT.

We adopt the unitary gauge, $\phi=\phi(t)$, and consider perturbations around the flat FLRW metric, $\D s^2=-\D t^2+a^2(t)\D \Vec{x}^2$.
The Horndeski action can then be expressed in terms of the lapse function $N$, the shift vector $N_i$, the determinant of the three-dimensional metric $h=\det h_{ij}$, and the extrinsic and intrinsic curvature tensors of constant-$t$ hypersurfaces, $K_{ij}$ and ${}^{(3)}R_{ij}$, as follows:
    \begin{align}
    S=\int \D^4x\sqrt{h}\frac{M^2}{2}\Big[&-(1+\delta N)\delta \mathcal{K}_2+(1+\alpha_T){}^{(3)}R+H^2\alpha_K\delta N^2-2H\alpha_{B}\delta K\delta N+{}^{(3)}R\delta N+\alpha_{V}\delta N\delta\mathcal{K}_2\Big],
    \label{eq:EFTactionU}
\end{align}
where $H:=\dot{a}/a$, $\delta N:=N-1$, $\delta K^j_i:= K^j_i-H\delta^j_i$, and $\delta \mathcal{K}_2:=\delta K^2-\delta K^j_i\delta K^i_j$.
We have kept higher-order terms that are responsible for the Vainshtein mechanism effective on small scales.
The relation between the coefficients in the above action and the functions appearing in the Horndeski action is given by\footnote{Different conventions are adopted in the literature for the definition of $\alpha_B$ and the sign in front of $G_3$ in the action.
Our convention coincides with that in Refs.~\cite{Bellini:2014fua,Zumalacarregui:2016pph}, but our $\alpha_B$ differs by a factor of $-2$ from that in Ref.~\cite{Dima:2017pwp}.}
\begin{align}
    M^2&=2(G_4-2XG_{4X}),\\
    \alpha_M&=\frac{1}{H}\frac{\D \ln M}{\D t},\\
    M^2H^2\alpha_K&=2X(K_X+2XK_{XX}-2G_{3\phi}-2XG_{3\phi X})+12\dot{\phi}XH(G_{3X}+XG_{3XX}-3G_{4\phi X}-2XG_{4\phi XX})\notag \\ 
    &\quad 
    +12XH^2(G_{4X}+8XG_{4XX}+4X^2G_{4XXX}),\\
    M^2H\alpha_B&=-2M^2H\alpha_V+2\dot{\phi}(XG_{3X}-G_{4\phi}-2XG_{4\phi X}),\\
    M^2\alpha_T&= 4XG_{4X},\\
    M^2\alpha_V&=-4X\left(G_{4X}+2XG_{4XX}\right).
\end{align}

The physical implications of these coefficients are as follows: $\alpha_M$ measures the time dependence of the effective Planck mass $M$,
$\alpha_B$ quantifies the kinetic mixing between the scalar and tensor sectors,
$\alpha_T$ characterizes the deviation in the propagation speed of gravitational waves relative to that of photons,
and $\alpha_V$ indicates the onset of modifications that appear at cubic order in perturbations.

\section{Peculiar velocity of the shell}
\label{app:velocity}

In this appendix, we derive the peculiar velocity $\vec{v}$ in the linear regime to see the appropriate initial conditions for a shell.
In the linear perturbation theory, the continuity equation reads
\begin{align}
  \frac{\pd\delta_L}{\pd t}+\frac{1}{a}\Vec{\cd}\cdot\vec{v}=0.
\end{align}
where we now use the comoving coordinates.
As in the standard cosmological model based on general relativity, the evolution equation for $\delta_L$ does not contain spatial derivatives acting on $\delta_L$ (see Eq.~\eqref{eq:linear}).
This allows us to write
\begin{align}
  \delta_L(t,\Vec{x})=D(t)\widetilde\delta_L(\vec{x}),
\end{align}
where the time-dependent part $D(t)$ is called the linear growth factor.
It is then easy to see that
\begin{align}
    \frac{\pd \delta_L(t,\vec{x})}{\pd t}
    =fH\delta_L(t,\vec{x}),
\end{align}
where we have defined the linear growth rate $f(t)$ as
\begin{align}
  f(t):=\frac{\D \ln D}{\D \ln a}.
\end{align}
The continuity equation now reads
\begin{align}
  \Vec{\cd}\cdot\vec{v}=-faH\delta_L(t,\vec{x}).
\end{align}
This equation can be integrated over a volume $V$ to obtain
\begin{align}
  \int_V\D V\Vec{\cd}\cdot\vec{v}
  =-faH\int_V\D^3x\,\delta_L(t,\vec{x}).
\end{align}
Using the Gauss theorem, we can write the left-hand side as the surface integral over the surface $S$ of the volume $V$.
Choosing $V$ as a sphere with comoving radius $r$ and $S$ as the surface of the sphere $S=4\pi r^2$, we can write
\begin{align}
  \oint_S\vec{v}\cdot \D \vec{S}=-faH\frac{4\pi r^3}{3}\Delta(t,r),
  \label{eq:int:surface:B}
\end{align}
where we have defined the average density contrast $\Delta(t,r)$ by
\begin{align}
  \Delta(t,r):=\frac{3}{4\pi r^3}\int_V\D^3x\delta(t,\vec{x}).
  \label{eq:ave:d}
\end{align}
In the spherically symmetric case, the left-hand side of Eq.~\eqref{eq:int:surface:B} reduces to $4\pi r^2 v_r(t,r)$, where $v_r$ is the radial component of $\Vec{v}$.
Thus,
\begin{align}
    v_r(t,r)=-\frac{1}{3}farH\Delta(t,r)=-\frac{1}{3}fRH\Delta(t,r),
\label{peculiarvelocity}
\end{align}
where $R$ is a physical radial coordinate $R(t)=a(t)r$.
If the initial condition is set at some moment $t=t_i$ in the deep matter-dominated era, one may use $f(t_i)\simeq 1$ to obtain
\begin{align}
    v_r(t_i,R_i)\simeq-\frac{1}{3}H(t_i)R_i\Delta(t_i,r_i).
    \label{eq:vini}
\end{align}
This is the peculiar velocity of a shell, and 
we use the sum of this and the Hubble flow as the initial velocity of the shell in Eq.~\eqref{eq:Rini}.

\section{Basic equation for the spherical collapse model}

To compute the void size function in Sec.~\ref{sec:vsf}, we also need to evaluate the critical density contrast $\delta_{\mathrm{c}}$ for halo formation.
For this purpose, we employ the spherical collapse model, which is widely used to estimate $\delta_{\mathrm{c}}$ semi-analytically~\cite{Bellini:2012qn,Barreira:2013xea,Albuquerque:2024hwv,Takadera:2025oyf}.
In the following, we present the basic equations governing the evolution of a top-hat matter overdensity.

The continuity and Euler equations for minimally coupled pressureless matter are given by
\begin{align}
    \dot\delta + \frac{1}{a}\Vec{\cd}\cdot\left[(1+\delta)\Vec{v}\right]
    &=0,\label{eq:con}
    \\ 
    \dot{\Vec{v}}+H\Vec{v}+\frac{1}{a}(\Vec{v}\cdot\Vec{\cd})\Vec{v}
    &=-\frac{1}{a}\cd\Phi,\label{eq:Euler}
\end{align}
where we now use the comoving coordinates and $\Vec{v}$ is the peculiar velocity field.
In the case of a spherical top-hat overdensity, we have
\begin{align}
    \delta(t,\Vec{x})=\delta(t)\quad (r=|\Vec{x}|\le r_*),
\end{align}
where $r_*$ is the comoving size of the overdense region.
The velocity field can be written as 
\begin{align}
    \Vec{v}(t,\Vec{x})=\frac{1}{3}\vartheta(t)\Vec{x},
\end{align}
where $\vartheta=\Vec{\nabla}\cdot\Vec{v}$.
Substituting these expressions into Eqs.~\eqref{eq:con} and~\eqref{eq:Euler},
we obtain
\begin{align}
    \dot\delta+\frac{1}{a}(1+\delta)\vartheta&=0,
    \\ 
    \dot\vartheta+H\vartheta+\frac{\vartheta^2}{3a}&=-\frac{1}{a}\nabla^2\Phi.
\end{align}
These equations are combined to give the nonlinear evolution equation for $\delta$:
\begin{align}
    \ddot\delta+2H\dot\delta-\frac{4}{3}\frac{\dot\delta^2}{1+\delta} 
    =(1+\delta)\frac{\nabla^2\Phi}{a^2}.
\end{align}
It is more convenient to write the right-hand side in terms of the first derivative of $\Phi$ because our approach directly relates $\Phi'$ with $\delta$.
Since all the terms in the left-hand side depend on only time, $\cd^2\Phi=r^{-2}(r^2\Phi')'$ must be independent of $r$, where a prime denotes differentiation with respect to $r$.
Imposing the regularity of $\Phi$ at $r=0$, it then follows that $\Phi$ must be of the form $\Phi(t,r)=\mcal Y(t)r^2$, and hence $\cd^2\Phi=3\Phi'/r=6\mcal Y(t)$.
Thus, we arrive at
\begin{align}
    \ddot\delta+2H\dot\delta-\frac{4}{3}\frac{\dot\delta^2}{1+\delta} 
    =3(1+\delta)\frac{\Phi'}{a^2r}.
\end{align}
To relate $\Phi'$ with $\delta$, we use Eq.~\eqref{eq:DPhi-DR} with $\delta m=\Omega_{\mrm{m}}\delta /2$ for a top-hat overdensity, yielding
\begin{align}
    \ddot\delta+2H\dot\delta-\frac{4}{3}\frac{\dot\delta^2}{1+\delta} 
    =3H^2(1+\delta)\left[
        \frac{\Omega_{\mrm{m}}}{2}\delta -\frac{1}{4}
        \left(\alpha_*-\sqrt{\alpha_*^2+\alpha_B\Omega_{\mrm{m}}\delta}\right)
    \right].\label{eq:delta}
\end{align}
This equation governs the nonlinear evolution of $\delta$.
We solve this equation to compute the evolution of overdensities in Sec.~\ref{sec:overdensity}.

\section{Excursion set theory and Sheth--van de Weygaert void size function}
\label{app:excursion}

In this appendix, we briefly review the excursion set theory~\cite{Bond:1990iw} and derive the Sheth--van de Weygaert void size function~\cite{Sheth:2003py} following Ref.~\cite{Zentner:2006vw}.

We begin by introducing the smoothed density contrast field,
\begin{align}
\delta(\vec{x},R_W)=\int \D^3 y\, W(|\vec{x}-\vec{y}|,R_W)\delta(\vec{y}),
\end{align}
where $W(|\vec{x}-\vec{y}|,R_W)$ is a window function with a smoothing scale $R_W$, and $\delta(\vec{x}):=[\rho(\vec{x})-\bar{\rho}]/\bar{\rho}$ is the density contrast field, with $\rho(\vec{x})$ and $\bar{\rho}$ denoting the local and mean matter densities, respectively.
We adopt the sharp-$k$ filter defined in Fourier space as
\begin{align}
  W(k,R_W)=\begin{cases}
    1 & (k\leq R_W^{-1}) \\
    0 & (k> R_W^{-1})
  \end{cases},
\end{align}
which, in real space, is written as
\begin{align}
  W(x,R_W)=\frac{1}{2\pi^2 R_W^3}\frac{\left[\sin(xR_W^{-1})-xR_W^{-1}\cos(xR_W^{-1})\right]}{(xR_W^{-1})^3}.
\end{align}
The Fourier transform of the smoothed density contrast is given by
\begin{align}
  \delta(\vec{k},R_W)=W(k,R_W)\delta(\vec{k}),
  \label{eq:delta_k}
\end{align}
where $\delta(\vec{k})$ is the Fourier transform of $\delta(\vec{x})$.
The variance of the smoothed density contrast field is given by
\begin{align}
  \sigma^2(R)=\frac{1}{2\pi^2}\int \D k k^2 P(k)W^2(k,R),
\end{align}
where $P(k)$ is the linear matter power spectrum.
In what follows we write $S(R)=\sigma^2(R)$.
Assuming that $\delta(\vec{x})$ is a Gaussian random field, the smoothed field $\delta(\vec{x},R)$ is also Gaussian random, with the probability distribution
\begin{align}
  \textsf{P}(\delta;R)\D\delta=\frac{1}{\sqrt{2\pi S(R)}}\exp\left[-\frac{\delta^2}{2S(R)}\right]\D\delta.
  \label{eq:pdf_delta}
\end{align}
Note that $S(R)$ is a monotonically decreasing function of the smoothing scale $R$.
This allows us to use $R$ and $S$ interchangeably and regard a function of $R$ as a function of $S$.

In the Press--Schechter formalism, 
one considers the cumulative probability for $\delta(\Vec{x},R)$ to be greater than some threshold $\delta_{\mathrm{c}}$,
\begin{align}
    F(R)=1-\int_{\delta_{\mrm{c}}}^\infty \mathsf{P}(\delta;R)\D\delta,
\end{align}
which amounts to the fractional volume occupied by collapsed objects larger than $R$.
In the limit $R\to 0$ ($S\to \infty$), one has $F=1/2$, which is half the expected value.
This underestimate is associated with the cloud-in-cloud problem:
$\delta(\vec{x},R)$ may be below $\delta_{\mathrm c}$ on some scale $R$, while exceeding it on a larger scale $R'>R$.
The excursion set theory, which is an extension of the Press--Schechter formalism, resolves this problem by identifying the largest smoothing scale $R$ at which $\delta(\vec{x},R)$ exceeds the threshold $\delta_{\mathrm{c}}$.

The essential concept for the excursion set theory is to regard $S$ as a ``time'' variable and consider an ensemble of ``trajectories'' $\delta(S)$.
Finding the largest $R$ at which $\delta(\Vec{x},R)$ exceeds $\delta_{\mathrm c}$ is equivalent to finding the ``time'' $S$ at which $\delta(\Vec{x},S)$ first crosses $\delta_{\mathrm c}$.
Along this line of thought, let us discuss the probability distribution $\Pi(\delta, S)$ of reaching $\delta$ at the time $S$, starting from some initial value at some initial moment (say, $\delta=0$ at $S=0$).
To derive the partial differential equation that $\Pi(\delta,S)$ obeys, we consider a trajectory passing through $\delta_1$ at time $S_1$ and arriving at $\delta_2=\delta_1+\Delta\delta$ at the next step at time $S_2=S_1+\Delta S$.
In the case of the sharp-$k$ filter, one can show that the transition probability from $(S_1,\delta_1)$ to $(S_1+\Delta S,\delta_1+\Delta \delta)$ is Gaussian with zero mean and variance $\Delta S$,
\begin{align}
  \Psi(\Delta\delta;\Delta S)=\frac{1}{\sqrt{2\pi \Delta S}}\exp\left[-\frac{(\Delta\delta)^2}{2\Delta S}\right],
  \label{eq:pdf_Delta}
\end{align}
which is independent of $(S_1,\delta_1)$.
By definition, we have
\begin{align}
    \Pi(\delta_2,S_2) 
    = \int_{-\infty}^{\infty}
    \Psi(\Delta\delta;\Delta S)\Pi(\delta_1,S_1)\D\delta_1
    .
\end{align}
Setting $S_1=S$ and $\delta_2=\delta$ and changing the integration variable from $\delta_1$ to $\Delta \delta=\delta_2-\delta_1$, we can rewrite this equation as
\begin{align}
  \Pi(\delta,S+\Delta S)=\int_{-\infty}^{\infty} \D(\Delta\delta) \Psi(\Delta\delta;\Delta S)\Pi(\delta-\Delta\delta,S),
\end{align}
which is called the Chapman--Kolmogorov equation.
Expanding this equation to first order in $\Delta S$ and to second order in $\Delta\delta$, we obtain
\begin{align}
  \Pi(\delta,S)+\frac{\pd \Pi(\delta,S)}{\pd S}\Delta S
  =\int \D(\Delta\delta) \Psi(\Delta\delta;\Delta S)\left[\Pi(\delta,S)-\frac{\pd \Pi(\delta,S)}{\pd \delta}\Delta\delta+\frac{1}{2}\frac{\pd^2 \Pi(\delta,S)}{\pd \delta^2}(\Delta\delta)^2\right].
\end{align}
Using $\ev{\Delta\delta}=0$ and $\ev{(\Delta\delta)^2}=\Delta S$, we can perform the integration over $\Delta\delta$ to get
\begin{align}
  \frac{\pd\Pi}{\pd S}=\frac{1}{2}\frac{\pd^2 \Pi}{\pd \delta^2}.
  \label{eq:Fokker-Planck}
\end{align}
Thus, $\Pi(\delta,S)$ obeys the Fokker--Planck equation for the probability distribution of a random walk, with the ``time'' variable $S$.
One can compute the void size function as well as the halo mass function by solving Eq.~\eqref{eq:Fokker-Planck} with appropriate boundary conditions.

In the case of halos, the cloud-in-cloud problem is resolved by imposing an absorbing boundary at $\delta=\delta_{\mathrm c}$, where $\delta_{\mathrm c}$ is the critical density contrast computed by the use of the spherical collapse model.
In the case of voids, an underdense region may be embedded in a collapsing overdense region.
This is the void-in-cloud problem.
In this case, we must introduce two absorbing barriers at $\delta=\delta_{\mathrm c}$ and $\delta=\delta_{\mathrm{sc}}$.
Thus,
we impose the boundary conditions $\Pi(\delta_{\mathrm c},S)=0$ and $\Pi(\delta_{\mathrm{sc}},S)=0$, and solve Eq.~\eqref{eq:Fokker-Planck}
with the initial condition $\Pi(\delta,0)=\delta_{\mathrm D}(\delta)$,
where $\delta_{\mrm{D}}$ is the Dirac delta function.
The solution is given by
\begin{align}
    \Pi(\delta,S)=\frac{2}{\delta_{\mrm{c}}-\delta_{\mrm{sc}}}\sum_{n=1}^{\infty}\left\{\cos\left(\frac{n\pi}{\delta_{\mrm{c}}-\delta_{\mrm{sc}}}\delta\right)-\cos\left[\frac{n\pi}{\delta_{\mrm{c}}-\delta_{\mrm{sc}}}(2\delta_{\mrm{sc}}-\delta)\right]\right\}\exp[-\frac{n^2\pi^2}{2(\delta_{\mrm{c}}-\delta_{\mrm{sc}})}S].
\end{align}
The fraction of trajectories that have penetrated either barrier by the time $S$ is
\begin{align}
      F(S)=1-\int_{\delta_{\mrm{sc}}}^{\delta_{\mrm{c}}}\Pi(\delta,S)\D \delta.
\end{align}
The differential probability for a first crossing 
is then
\begin{align}
   \frac{\D F}{\D S}\D S=-\frac{1}{2}\left.\frac{\pd\Pi}{\pd\delta}\right|^{\delta_{\mrm{c}}}_{\delta_{\mrm{sc}}}\D S,
\end{align}
but only the lower boundary is of our interest because
the trajectories piercing it
correspond to the void regions.
Therefore, we take
\begin{align}
  f_{\ln S}\D S&=\frac{1}{2}\left.\frac{\pd \Pi}{\pd\delta}\right|_{\delta_{\mrm{sc}}}\D S\notag\\
  &=\sum_{n=1}^{\infty}n\pi x^2\sin(n\pi\mathcal{D})\exp[-\frac{(n\pi x)^2}{2}]\D S,
  \label{app:eq:dprob}
\end{align}
where
\begin{align}
  \mathcal{D}:=\frac{|\delta_{\mrm{sc}}|}{\delta_{\mrm{c}}+|\delta_{\mrm{sc}}|},\qquad x:=\frac{\mathcal{D}}{|\delta_{\mrm{sc}}|}\sigma.
\end{align}
This result was derived in Ref.~\cite{Sheth:2003py}.
In the main text, we use $\D F/\D \sigma=2\D F/\D S$ rather than $\D F/\D S$ as the differential probability, and thus~\cite{Jennings:2013nsa}
\begin{align}
  f_{\ln \sigma}
  =2\sum_{n=1}^{\infty}n\pi x^2\sin(n\pi\mathcal{D})\exp[-\frac{(n\pi x)^2}{2}].
\end{align}

\bibliography{refs}
\bibliographystyle{JHEP}
\end{document}